\documentclass[12pt,square,twocolumn,twoside,lineno]{tem}
\usepackage[utf8]{inputenc}

\runningtitle{Dynamic Reconfiguration of Brian Functional Network in Stroke} 
\runningauthor{Kaichao Wu \textit{et al.}}

\title{Dynamic Reconfiguration of Brian Functional Network in Stroke}

\author[1,2]{Kaichao Wu}
\author[3]{Beth Jelfs}
\author[2]{Katrina Neville}
\author[4,*]{Wenzhen He}
\author[1,*]{John Q.Fang}

\affil[1]{Department of Biomedical Engineering, College of Engineering, Shantou University, Shantou, P.R.China}
\affil[2]{School of Engineering, RMIT University, Melbourne, Australia}
\affil[3]{Department of Electronic, Electrical and Systems Engineering, The University of Birmingham, Birmingham, UK.}
\affil[4]{1st Affiliated Hospital of Shantou University Medical College, P.R.China.}    

\begin{abstract}
The brain continually reorganizes its functional network to adapt to post-stroke functional impairments. Previous studies using static modularity analysis have presented global-level behavior patterns of this network reorganization. However, it is far from understood how the brain reconfigures its functional network dynamically following a stroke. This study collected resting-state functional MRI data from 15 stroke patients, with mild (n = 6) and severe (n = 9) two subgroups based on their clinical symptoms. Additionally, 15 age-matched healthy subjects were considered as controls. By applying a multilayer network method, a dynamic modular structure was recognized based on a time-resolved function network. Then dynamic network measurements (recruitment, integration, and flexibility) were calculated to characterize the dynamic reconfiguration of post-stroke brain functional networks, hence, to reveal the neural functional rebuilding process. It was found from this investigation that severe patients tended to have reduced recruitment and increased between-network integration, while mild patients exhibited low network flexibility and less network integration. It’s also noted that this severity-dependent alteration in network interaction was not able to be revealed by previous studies using static methods. Clinically, the obtained knowledge of the diverse patterns of dynamic adjustment in brain functional networks observed from the brain signal could help understand the underlying mechanism of the motor, speech, and cognitive functional impairments caused by stroke attacks. The proposed method not only could be used to evaluate patients' current brain status but also has the potential to provide insights into prognosis analysis and prediction.
\end{abstract}


\begin{document}

\maketitle

\section{Introduction}

Stroke is a common neurological disorder that can lead to significant impairment of cognitive and motor functions. However, due to brain plasticity, the stroke brain can adjust its network architecture to adapt to structural damage and compensate for the lost functions~\cite{RN144, RN16}. The brain functions are fulfilled by a set of functionally specialized modules, i.e., distributed brain regions that interact and cooperate with each other, either within modules or between modules, in response to the functional demands of the external environment~\cite{RN27, RN22}. Therefore, the altered functional network of the stroke brain implies that it is reconfiguring its modular structure to support the post-stroke plasticity~\cite{RN28}. 

In this regard, functional neuroimaging data, particularly resting-state functional MRI, have contributed enormously to understanding the reorganization mechanisms of network modules underpinning post-stroke plasticity and brain adaptability~\cite{RN31, RN30, RN29}. A frequent observation is the reduction of modularity between sub-networks after a stroke~\cite{RN19, RN17, RN18}. 
This reduction in modularity reflects the decreased segregation between different functional domains and integration within domains; to some extent explaining the post-stroke clinical deficits~\cite{RN9, RN18}. The reduced modularity usually lasts a few weeks to a month after the stroke, following which the modular brain network gradually recovers, in parallel with the functional improvement (e.g., improvements language~\cite{RN19} and attention~\cite{RN12}). 

Nevertheless, of note that these findings typically rest on a static representation or a single brain network built from an entire resting-state functional MRI scan. While a static construction is valuable and useful, a growing body of literature on time-varying networks suggests that the temporal dynamics of the modular brain should be assessed~\cite{RN6, RN32}. In addition, the time dependence of the modularity recovery implies that the dynamic reconfiguration of the brain networks could be the root source of decreased or increased static modularity, thus further emphasizing the need for evaluation of post-stroke module dynamics. Furthermore, the brain's dynamic reconfiguration has been proven to be a promising avenue for creating novel biomarkers of diseases, such as attention-deficit/hyperactivity disorder~\cite{RN23} schizophrenia~\cite{RN24}, temporal lobe epilepsy~\cite{RN26} and depression~\cite{RN274}. However, if and how the brain network dynamically reconfigures itself following a focal stroke, particularly under different levels of clinical severity, remains not fully understood.

\begin{figure*}[!t]
    \centering
    \includegraphics[width=0.9\linewidth]{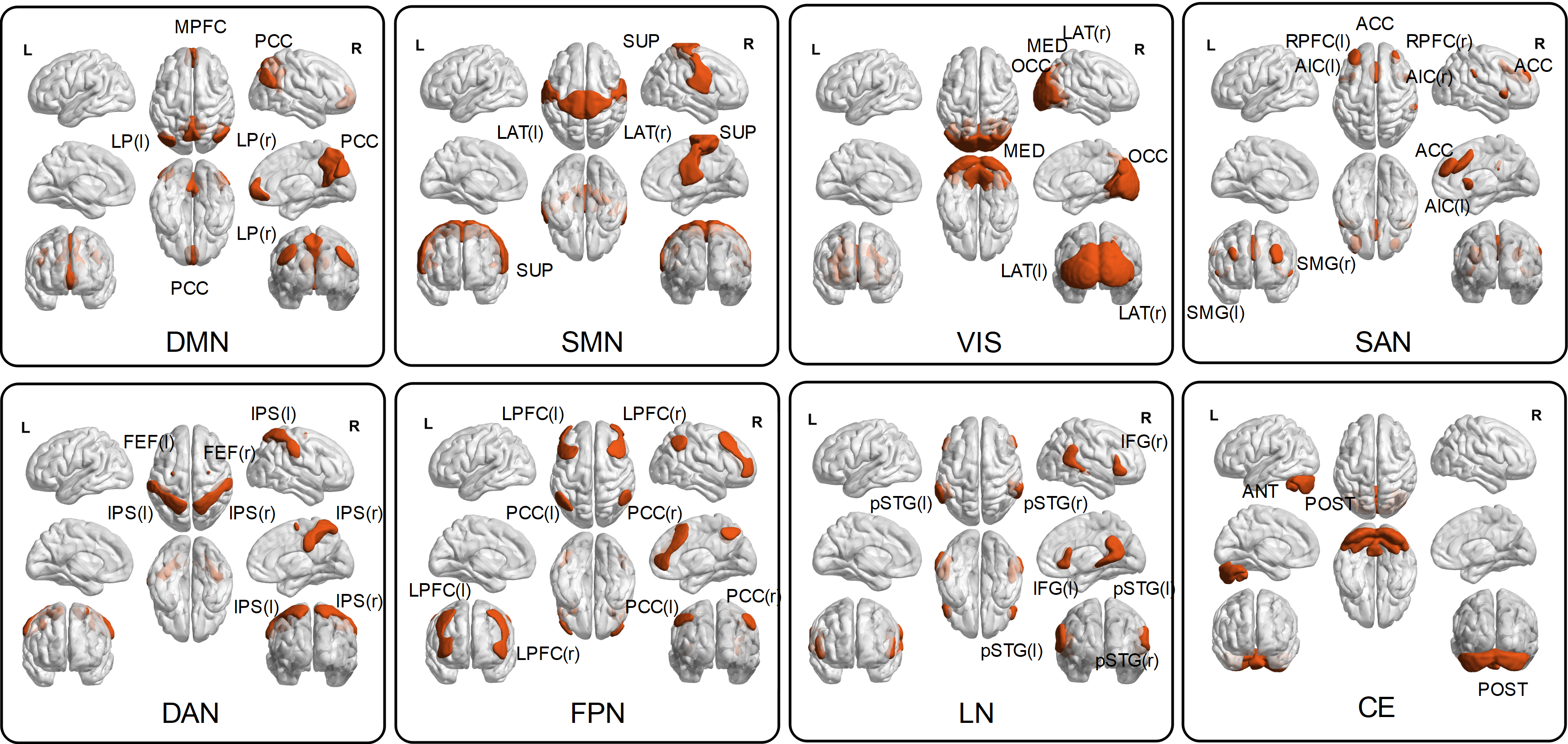}
    \caption{Spatial mapping of predefined 32 ROIs and 8 corresponding networks for multilayer dynamic analysis. Default mode network (DMN): the medial prefrontal cortex (MPFC), precuneus cortex (PCC), bilateral lateral parietal (LP); sensorimotor network (SMN): superior, bilateral lateral; visual network (VIS): medial, occipital, bilateral lateral; salience network (SAN): anterior cingulate cortex (ACC), bilateral anterior insula (AI), rostral prefrontal cortex (RPFC), and supramarginal gyrus (SMG); dorsal attention network (DAN): bilateral frontal eye field (FEF) and intraparietal sulcus (IPS); fronto-parietal network (FPN): the bilateral lateral prefrontal cortex (LPFC) and posterior parietal cortex (PPC); language network (LN): bilateral inferior frontal gyrus (IFG) and posterior superior temporal gyrus (pSTG); and cerebellar network (CE): anterior, posterior.\label{fig:rois}}
\end{figure*}

Therefore, this study investigated the dynamic reconfiguration of functional brain networks in stroke patients with different degrees of clinical symptoms. Specifically, the fMRI data from 15 stroke subjects with two degrees of severity (mild: 6 and severe: 9) and 15 age-matched healthy controls were analysed with a multilayer network model. We hypothesize that the post-stroke brain dynamic reconfigures its functional network according to clinical severity. The hypothesis is twofold: first, the brain functional network of stroke patients undergoes dynamic changes. If these changes do happen, they can be evidenced by highly significant alterations in the measurements that characterize dynamic reconfiguration following a stroke. Second, the dynamic behaviours of the brain functional network exhibit highly significant alteration between subgroups, i.e., reconfiguration pattern differs between patients with distinct degrees of clinical symptoms.  

\section{Materials and Methods}
\subsection{Participants}
The stroke samples examined in this study were from fifteen ischemic stroke patients admitted to the 1st affiliated hospital of Shantou University Medical College (SUMC, mean age 63.8 years with a standard deviation of 11.68 years, 4 male/11 females, mean day of MRI scan post-stroke is 23.06 with a standard deviation of 4.32). The patients were recruited from a greater study approved by the medical research ethics committees of the named hospitals, and all participants signed informed consent. All participants were right-handed, had normal vision, and had no hearing deficits. The patient inclusion criteria and the details of recruited 15 patients can be seen in the Supplementary Material (a-1 and a-2).

Patients with National Institutes of Health Stroke Scale (NIHSS)$>$5 were assigned to a severe subgroup; otherwise, they were assigned to the mild subgroup~\cite{RN260}. In addition, fifteen age-matched healthy samples from our previous research served as control groups~\cite{wu2023tracking} (7 male/8 female, mean age 68.6 years with a standard deviation of 6.4 years). The demographic characteristics of all participants and the clinical features of stroke patients can be seen in the Supplementary Material (a-3).

\subsection{MRI Acquisition}
Acquisition of MRI data was all performed on a Discovery standard 3.0 T scanner using an 8-channel head coil at the MRI center of SUMC. The high-resolution T1 anatomical images were acquired with a multi-planar rapidly acquired gradient echo sequence with 1 mm isotropic voxels, a $256\times256$ matrix size, and a 9-degree flip angle (129 slices, repetition time (TR) = 2250 ms, Time of echo (TE) = 4.52 ms). With the T1, the lesion profile of all patients has been created as a lesion overlap map (the details and corresponding lesion map can be seen in the Supplementary Material b-1). 

Resting-state functional MRI was collected after the anatomical scan using single-shot gradient-echo EPI sequence:  TR = 2,000 ms; TE = 30 ms; flip angle = 90; field of view = $240\times240$ mm$^2$; matrix size = $64 \times 64$; number of slices = 25; and voxel size = $3.43\times 3.43 \times 5.0$ mm$^3$ with no gap; and 210 volumes acquired in 7 min.

\begin{figure*}[!t]
    \centering
\includegraphics[width=0.8\textwidth]{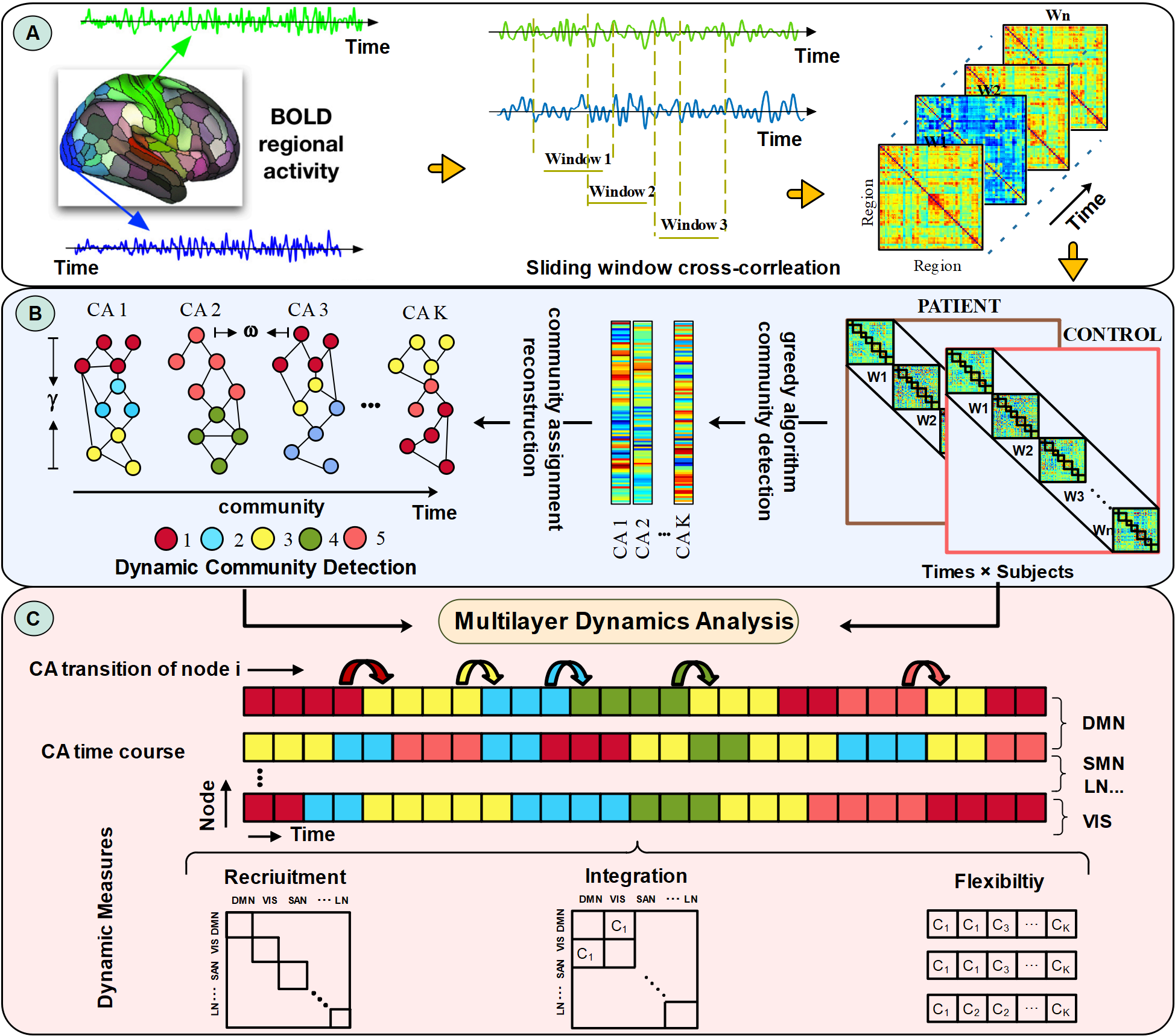}
    \caption{Flowchart of the multilayer dynamics analysis framework.}
    \label{fig:pipeline}
\end{figure*}

\subsection{FMRI Data Preprocessing and Head Motion Control}
The functional MRI scans were processed using a customized pre-processing pipeline in the CONN functional connectivity toolbox~\cite{RN_conn} in conjunction with the Statistical Parametric Mapping software package (SPM12)\cite{penny2011statistical}. 

For all subjects, the first 10 functional volumes were removed to obtain a steady blood oxygenation level-dependent activity signal. The remaining 200 images were corrected for slice timing and head motion, and they were normalized to Montreal Neurologic Institute (MNI) space. The non-smoothed functional images were fed into the default denoising pipeline for the elimination of confounding effects and temporal band-pass filtering. Preprocessing and denoising details and their quality assessments are available in the supplementary Material (b-2).

The head motion effect was controlled in functional connectivity analysis by calculating the individual framewise displacements (FD). Participants with a maximum displacement exceeding 1.5 mm and a maximum rotation above 1.5 degrees were excluded. In practice, no subjects exceeded these criteria so non were excluded. In addition, 24 motion parameters, calculated from the six original motion parameters, were regressed out as nuisance covariates. Finally, there was no significant group difference in mean FD when comparing the 15 stroke patients with the 15 healthy controls. Detailed results are available in the Supplemental Material (b-3).

\subsection{Functional Connectivity Estimation}
Functional connectivity is estimated by calculating the Person’s correlation coefficient between pairwise time series of spatially distinct brain regions. These regions are generated from anatomically or functionally parcelled brains, also known as brain parcels~\cite{RN23}. In this study, a functional brain parcellation provided by CONN was used to investigate the changing network configuration due to stroke lesions. This parcellation comprises 32 regions of interest (ROIs) which can cover the whole-brain area and be formed by eight large-scale networks/systems (details can be seen in \autoref{fig:rois}). Then, for $N$ ROIs, a $N\times N$ functional connectivity matrix $A$ can be created, where each entry $A_i$ is a pairwise Person’s coefficient between ROIs $i$ and $j$. To eliminate the bias, Fisher’s Z-transformation was applied to the functional connectivity matrices to obtain normally distributed Z-scores, and only the positive values were retained in the further connectivity analysis. The traditional static functional connectivity analysis for stroke patients is also available in the Supplemental Material (b-4).

\subsection{Static Modularity}
Static modularity is a theoretical graph metric measuring the segregation between distinct brain function systems~\cite{RN3}. As suggested in previous studies~\cite{RN17, RN139}, Newman’s method was implemented in the Brain Connectivity Toolbox for the static modularity calculation~\cite{RN258}. Modularity was calculated at edge densities ranging from 4 to 20\% with the symmetric treatment of negative weights consistent with reference~\cite{RN272, RN17, RN139}. The modularity calculated at each edge density was tested to see if there were significant differences between subgroups (mild vs. severe, severe vs. control, mild vs. control), and the average values across densities with a significant group effect were used as the final static modularity.

\subsection{Multilayer Modularity}

\textbf{Dynamic functional connectivity estimation}. Multilayer modularity calculation for dynamics analysis is based on dynamic functional network connectivity (dFNC) estimation. In this context, the common sliding window scheme was first employed to obtain temporal slices (\autoref{fig:pipeline}A). At each time point, a tapered window was used, which was obtained by convolving a rectangle (equal to the window size) with a Gaussian ($\sigma$ = 3). While the optimal choice of window width setting in the sliding window scheme is still under debate, prior studies have provided evidence that the number of communities fluctuates narrowly with a window width of 100s (50 TR)~\cite{RN22,liu2022dynamic}. In this paper,  the window width was opted for 50 TR and a step size of 1 TR~\cite{RN11}(The alternative option with  a window width of 20 TR has been shown in Supplemental Material (c-2)). For each sliding window, Fisher’s z-transform of Pearson’s correlation coefficient between all pairs of segmented timeseries was computed to estimate the dynamic functional connectivity. 

\textbf{Multilayer network detection}. using the estimated dFNC, a multilayer network can be detected as follows: First, the dynamical functional connectivity matrices of stroke patients and healthy controls were concatenated along the diagonal to produce their initial community profile (obtaining a matrix with $4,768\times4,768$, where $4,768 = 149\times32$, being the number of sliding windows and the size of each window respectively). Then, a Louvain-like greedy community detection algorithm was used for dynamic community detection. This algorithm optimizes the multilayer modularity partition by maximizing the modularity quality function~\cite{RN48}, which is defined as:
\begin{equation}
    Q_M=\frac{1}{2\mu}\sum_{ijlr}{\left[\left(A_{ijl}-\gamma \frac{k_{il}k_{jl}}{2m_l} \right) \delta _{lr}+\delta_{ij}\omega _{jlr}\right] \delta \left(g_{il},g_{jr}\right)},
\end{equation} 
where 
\begin{itemize}
    \item $A_{ijl}$ is the weight of the edges between nodes $i$ and $j$ at layer $l$;
    \item $k_{jl}$ is the weighted degree of node $j$ in layer $l$, that is the sum of the weights of the edges connected to node $j$ in layer $l$;
    \item $m_l$ is the total nodal weighted degrees in layer $l$;
    \item $\mu = \frac{1}{2}\sum_{ij}{(k_{jr}+c_{jr})}$ is the sum of the weights of the dynamic functional connectivity matrix;
    \item $c_{jr}=\sum_{l}{\omega_{jrl}}$ and $\omega_{jrl}$ is the edge strength between node $j$ in layer $l$ and node $j$ in layer $r$. 
    \item $\delta_{ij}$ denotes the Kronecker $\delta $-function, where $\delta _{ij}=1$ if $i=j$, otherwise $0$;
    \item $g_{il}$ and $g_{jr}$ represent the community node $i$ is assigned to in layer $l$ and node $j$ in layer $r$ respectively;
    \item $\delta \left( g_{il},g_{jr} \right)=1$ if $g_{ir}=g_{jl}$, otherwise $0$;
    \item The parameters $\gamma$ and $\omega$ are the intra-layer and inter-layer coupling parameters, controlling the number of modules detected in layers and across layers.
\end{itemize}
The final values of the two hyperparameters were determined with the grid-search method, the details of which can be seen in the Supplementary Material (c-1).

The final optimization associates the functional connectivity modularity to each sliding window. Hence, for the 149 sliding windows of each subject obtained in the multilayer resolution, there would be 149 community assignment (CA) vectors; the length of each CA vector is 32, corresponding to the number of predefined ROIs. After reconstruction, a multilayer network with a complex and rich community modularity structure spanning the time-varying layers can be obtained.

\begin{figure*}[htbp]
    \centering
    \includegraphics[width = 0.45\linewidth]{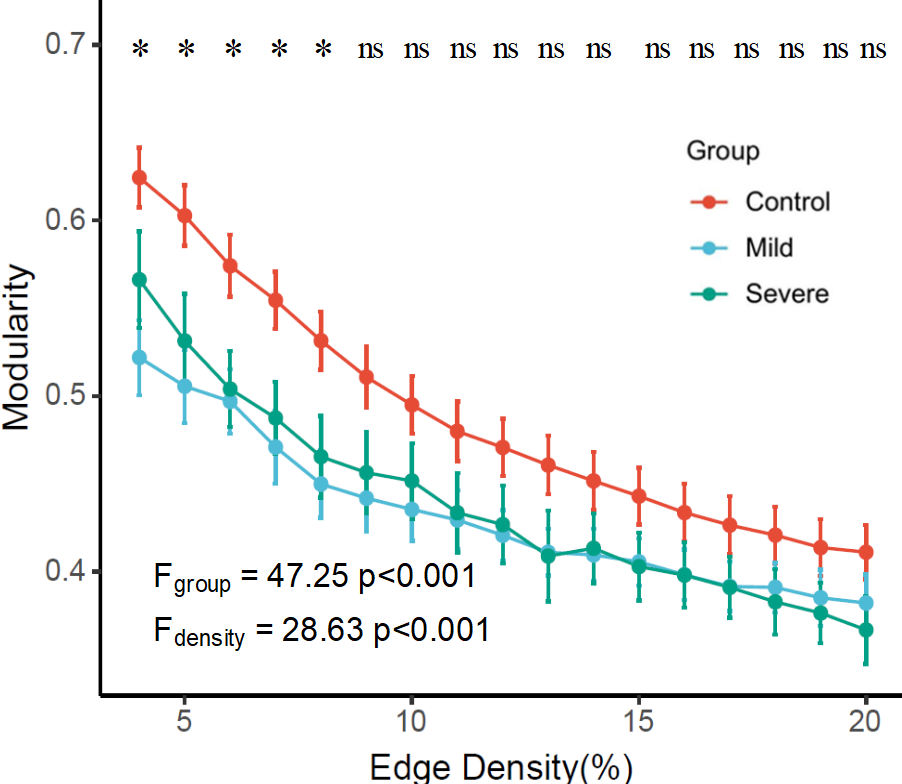}\hfill
    \includegraphics[width = 0.45\linewidth]{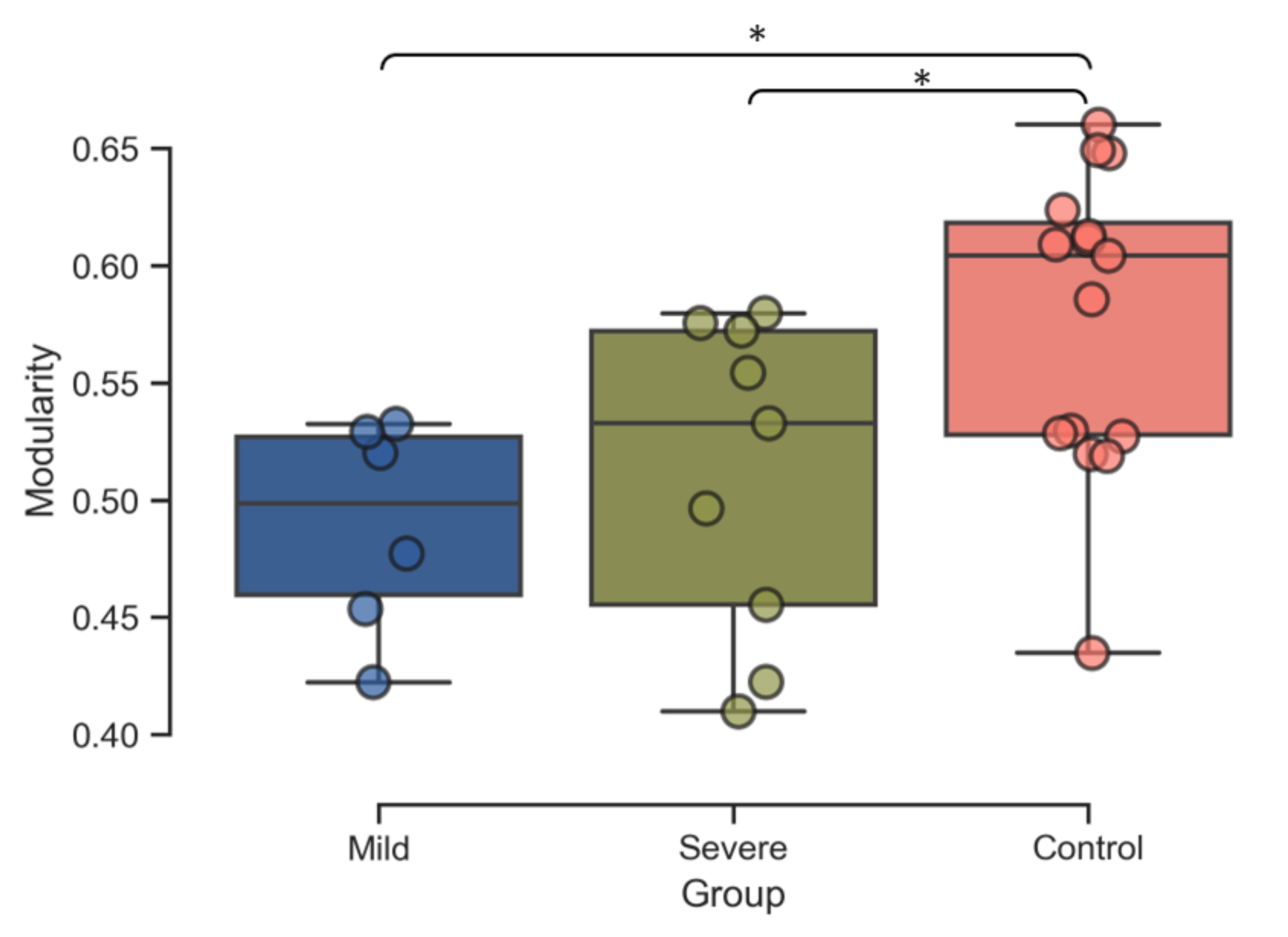}
    \caption{Left. Modularities across edge densities are shown for mild, severe, and healthy controls. Right. The final static modularity difference between different subgroups; horizontal lines indicate group means, and asterisks represent significant differences at $p< 0.05$ Bonferroni corrected, $ns$ denotes no significance. \label{fig:static}}
\end{figure*}

\textbf{Multilayer dynamics analysis.} For each participant from the two different subgroups, three measures: recruitment, integration and flexibility, were calculated to characterize the multilayer network dynamics based on the detected dynamic community structure.

Recruitment and integration quantify the dynamic functional interaction within and between brain functional systems. Precisely, recruitment is measured by the fraction of layers in which ROIs from the same functional system are assigned to the same community~\cite{RN17}. The recruitment of a given predefined functional system $S$ is defined as: 
\begin{equation}
    R_S=\frac{1}{n_S}\sum_{i\in S}{\sum_{j\in S}{P_{ij}}},
\end{equation}		
where $n_S$ is the number of ROIs belonging to the system $S$; $P_{ij}$ is the allegiance matrix of the multilayer networks, which is defined as $P_{ij}=\frac{1}{T}\sum_{t=1}^T{a_{ij}^{t}}$ with $a_{ij}^{t}=1$ if in layer $t$ nodes $i$ and $j$ are assigned to the same community, and 0 otherwise. Similar to recruitment, the integration of a given predefined functional system $S$ is defined as:
\begin{equation}
     I_S=\frac{1}{N-n_S}\sum_{i\in S}{\sum_{j\notin S}{P_{ij}}}.
\end{equation}

The system of interest is highly functionally integrated when its functional regions are frequently assigned to the same community as other regions. Therefore, to quantify this, an integration coefficient can also be defined between different functional systems~\cite{RN13}. The integration between functional system $S_k$ and $S_l$ is calculated as:
\begin{equation}
    I_{S_kS_l}=\frac{1}{n_{S_k}n_{S_l}}\sum_{i\in S_k}{\sum_{j\in S_l}{P_{ij}}}.
\end{equation}
The higher the between-system integration, the stronger the functional coordination between systems. This study investigated both within-system and between-system integration alterations caused by stroke lesions.

Flexibility characterizes the community stability of a system in multilayer resolution~\cite{RN18}. The flexibility of a system corresponds to the average number of times that its brain regions change module allegiance. The system $S$’s flexibility is defined as:
\begin{equation}
	     F_S=\frac{1}{n_s \times \left(T-1 \right)}\sum_{i\in S}{\sum_{t=1}^T{b_i}},
\end{equation}
where $n_S$ is the number of regions belonging to the system $S$, $T$ is the multilayer resolution, and $b_i=1$ if in the next layer $t+1$ the node $i$ is assigned to a different community.

Noting that random effects in the Louvain-like greedy community detection algorithm exist in multilayer community detection, the multilayer modularity optimization was run 100 times. The mean of the corresponding dynamic measures from the 100 repetitions served as their final values. Besides, a permutation approach~\cite{RN6} was used for the normalization of these dynamic measures. Specifically, a null distribution was created from 1000 randomly permuted multilayer function connectivity matrices. The recruitment, integration, and flexibility are then divided by the mean of the corresponding null distribution to obtain normalized values.

\subsection{Statistical Analysis}
 A 2-sample t-test (control covariates: age, sex, and FD ) was performed on the static functional connectivity and modularity to determine if there were functional network changes between patient groups and controls. In addition, a three-level one-way ANOVA (level of significance $p<0.05$) was performed to investigate if there were static modularity differences in healthy controls and mild and severe patients. In case of significant ANOVA results, post hoc t-tests (mild patients, severe patients and controls) were performed. Correction for multiple comparisons was always applied whenever testing more than one hypothesis simultaneously (false discovery rate (FDR) correction $p< 0.05$).

\begin{figure}[!t]
    \centering
    \includegraphics[width= \columnwidth]{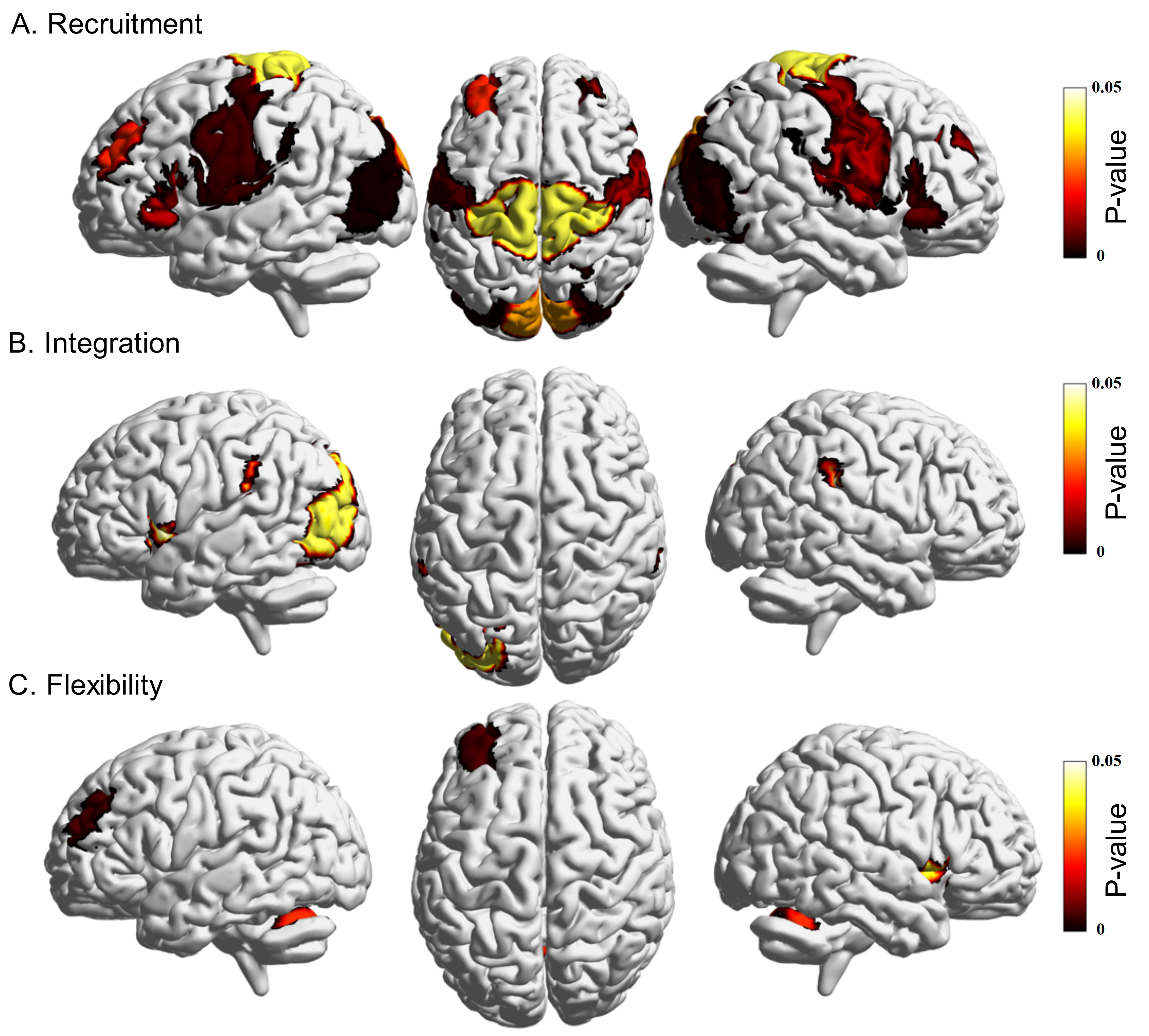}
    \caption{Brain regions exhibiting significant differences in ANOVA results for A. recruitment, B. integration, and C. flexibility.\label{fig：regions}}
\end{figure}

\begin{figure*}[t]
    \centering
    \includegraphics[width=0.9\textwidth]{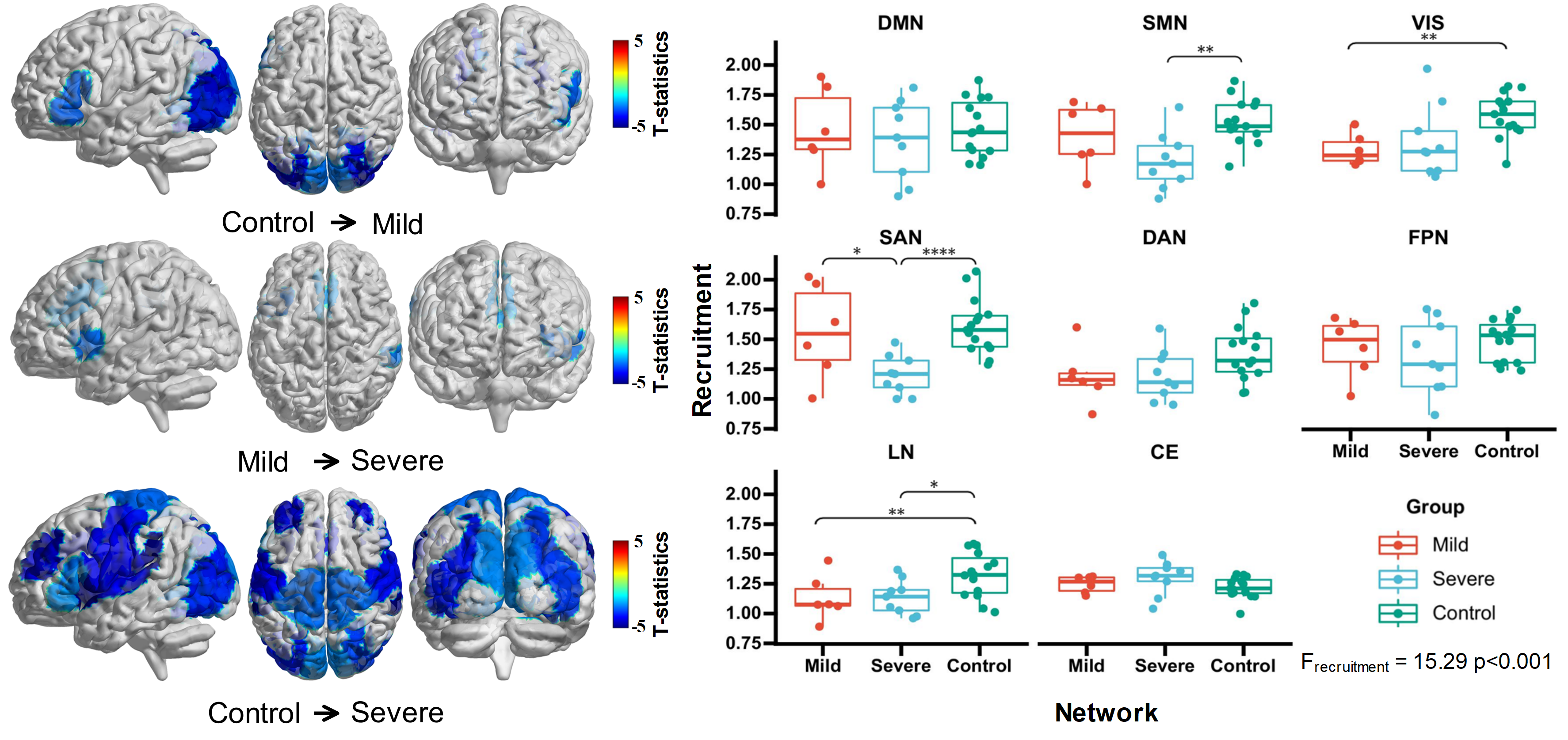}
    \caption{Results of post-hoc tests to examine the differences in recruitment between groups with distinct clinical symptoms.\label{fig：recrutiment}}
\end{figure*}

\section{Results}
\subsection{Whole Brain Static Modularity Alteration}
\autoref{fig:static} shows the static modularity for each group and each density. As expected, the edge density also has significant effects on the whole brain functional network modularity ($F(16) = 28.63$, $p< 0.0001$). Comparing the different edge densities, at only 5 densities, was the modularity significantly different between mild, severe, and control subgroups ($p_{0.04} = 0.011$, $p_{0.05}= 0.009$, $p_{0.06}= 0.014$, $p_{0.07} = 0.008$, $p_{0.08} = 0.014$). In general, the stroke patients had much lower functional network modularity ($F(2) = 47.25$, $p< 0.0001$), suggesting that the brain tends to have a less segregated functional network after stroke. The final static modularity also indicates that both the mild patients ($p = 0.02$, Bonferroni corrected) and the severe patients ($p = 0.04$, Bonferroni corrected) have lower network segregation than healthy controls. However, the significant effect on the static modularity was not detectable between the groups of mild and severe patients.

\subsection{Significant Brain Region Reconfiguration}
Static modularity reflects the average state that functional brain networks exhibit. However, the modular organization is not static but instead fluctuates constantly in response to the brain's functional demands, especially demands that have dramatic changes, such as when facing brain deficits. Three measures were produced to characterize this dynamic process based on the detected multilayer network. \autoref{fig：regions} shows the brain regions with significant differences in these measurements using three-level one-way ANOVA analysis results. According to the network parcellation, those brain regions with significant differences in recruitment are distributed across four networks: SMN, SAN, VIS, and LN (\autoref{fig：regions}A). The brain regions with significant differences in integration mainly reside in SAN and VIS (\autoref{fig：regions}B). Brain regions with significant differences in flexibility are distributed primarily in network SAN and CE (\autoref{fig：regions}C).

\begin{figure*}[tb]
    \centering
    \includegraphics[width=0.9\textwidth]{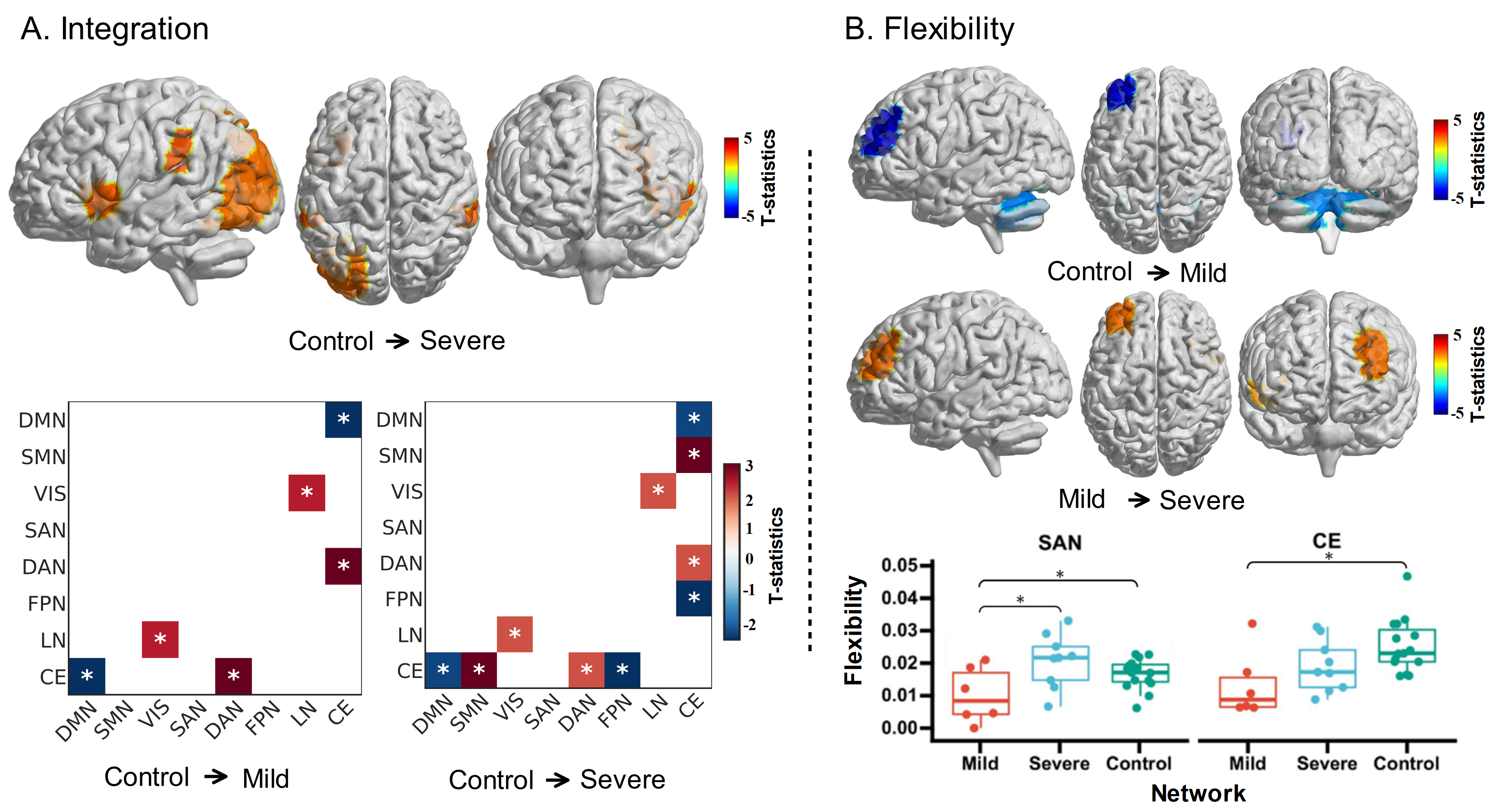}
    \caption {Post-hoc test results examining the differences in the \textbf{(A)} integration and \textbf{(B)} flexibility between groups with distinct clinical symptoms. \label{fig：integration}}
\end{figure*}

\subsection{Trends in Network Reconfiguration Based on Stroke Severity}
The brain regions which showed significantly different measurements between groups indicate that the brain networks reconfigure themselves after stroke. Next, we examined how dynamic reconfiguration is exhibited in brain functional networks, and whether these configuration patterns differ between patients with different stroke severity.

First, the between-group differences in recruitment are examined. The mild and severe patients show that most brain regions decline in recruitment compared to healthy controls. Severe patients exhibit more regions with declined recruitment compared to healthy controls than do the mild patients, implying that the number of regions with decreased recruitment increases as stroke severity grows. This inference was reinforced when solely comparing mild patients and severe patients, where the severe patients showed lower recruitment in ACC ($t= -2.225$, $p = 0.044$), left anterior insula ($t = 2.664$, $p = 0.019$) and right SMG ($t = -2.546$, $p = 0.024$) than the mild patients. \autoref{fig：recrutiment} illustrates the distribution of these regions with significant differences between subgroups. As nodal-level recruitment significantly differs, so the recruitment in large-scale systems exhibits differences (\textit{F} = 15.29 \textit{p} $<$ 0.0001). Post hoc comparison shows the mild patients had lower recruitment in VIS ($t = -4.973$, $p< 0.0001$) and LN ($t = -2.342$, $p= 0.030$), and the severe patients in SMN, SAN and LN compared to healthy controls (FDR corrected $p < 0.05$). 

Next, the group difference in integration between controls, mild patients, and severe patients was tested with the results shown in \autoref{fig：integration}. The post hoc comparison indicates that severe patients have higher integration in the anterior insula ($t = 2.762$, $p = 0.011$), right and left SMG (left: $t = 2.886$, $p = 0.009$. right: $t = 2.869$, $p = 0.009$), and lateral visual area ($t = 2.618$, $p = 0.016$) compared to controls. Neither mild patients and controls nor mild patients and severe patients differed in this aspect. In terms of integration between brain functional networks, the integration between DMN and CE ($F = 4.54$, $p = 0.019$), SMN and CE ($F = 3.66$, $p =0.039$), VIS and LN ($F = 4.18$, $p = 0.026$), DAN and CE ($F = 4.71$, $p = 0.017$), FPN and CE ($F = 3.45$, $p = 0.04$) was significantly altered. Post hoc t-tests, contrasting mild patients and healthy controls, revealed a stroke-induced decrease in integration between DMN and CE ($t = -2.124$, $p = 0.036$) but an increase between VIS and LN ($t = 2.208$, $p = 0.040$), DAN and CE ($t = 2.757$, $p = 0.013$). In contrast, severe patients comprised a decrease in integration between DMN and CE ($t = -2.653$, $p = 0.015$), FPN and CE ($t = -2.899$, $p = 0.008$), but an increase between SMN and CE ($t = 3.345$, $p = 0.003$), VIS and LN ($t = 2.213$, $p = 0.037$), and DAN and CE ($t = 2.218$, $p = 0.037$) when compared to healthy controls ($p<0.05$, FDR-corrected). Mild and severe patients did not feature significant differences in between-network integration after correction for multiple comparisons. \autoref{fig：integration}A. illustrates the details on the integration of altered network pairs.

Lastly, we investigated the between-group difference in flexibility. While a significant effect in flexibility was not detected when contrasting severe patients and healthy controls, mild patients featured significantly different flexibility in brain regions and functional networks compared to both controls and severe patients. In particular, the majority of the significantly altered areas resided in the salience and cerebellum functional domains. For example, mild patients have lower flexibility in the left RPFC than both other groups (to controls: $t = -4.078$, $p = 0.0006$; to severe patients: $t = -2.648$, $p = 0.020$). The lower flexibility in the mild patients was also exhibited in the anterior cerebellum when contrasting with controls ($t = -2.433$, $p = 0.025$) and in the right insula when contrasting with severe patients ($t = -2.039$, $p = 0.038$). A similar trend between groups is also observed in terms of functional network flexibility. Mild patients not only featured less flexibility in SAN than severe patients ($t = -2.410$, $p = 0.032$) but also lower flexibility in SAN ($t = -2.842$, $p = 0.010$) and CE ($t = -2.714$, $p = 0.035$) than controls. Notably, flexibility in SAN did not go down further as severity increased. When compared to mild patients, increased SAN flexibility was observed in severe patients. The box plot in \autoref{fig：integration}B shows the two networks (SAN and CE) with significantly different flexibility. Severe patients and controls did not differ in this regard.

Collectively, patients, no matter which level of severity, show remarkably consistent reduced recruitment compared to healthy controls. This reduction seems to exhibit continuity, as lower recruitment was observed in severe patients compared to mild patients. On top of that, the post-stroke dynamic reconfiguration can be represented by pairwise network integration instead of within-network integration. The mild and severe patients shared increased DMN-CE and decreased VIS-LN and DAN-CE integration. Regarding flexibility, this dynamic network measure has a significant group difference in SAN and CE. Of note is that SAN flexibility displays a U-shaped curve as severity rises, which exhibits a converse trend against SAN recruitment.

\begin{figure*}[!t]
    \centering
    \includegraphics[width=0.9\textwidth]{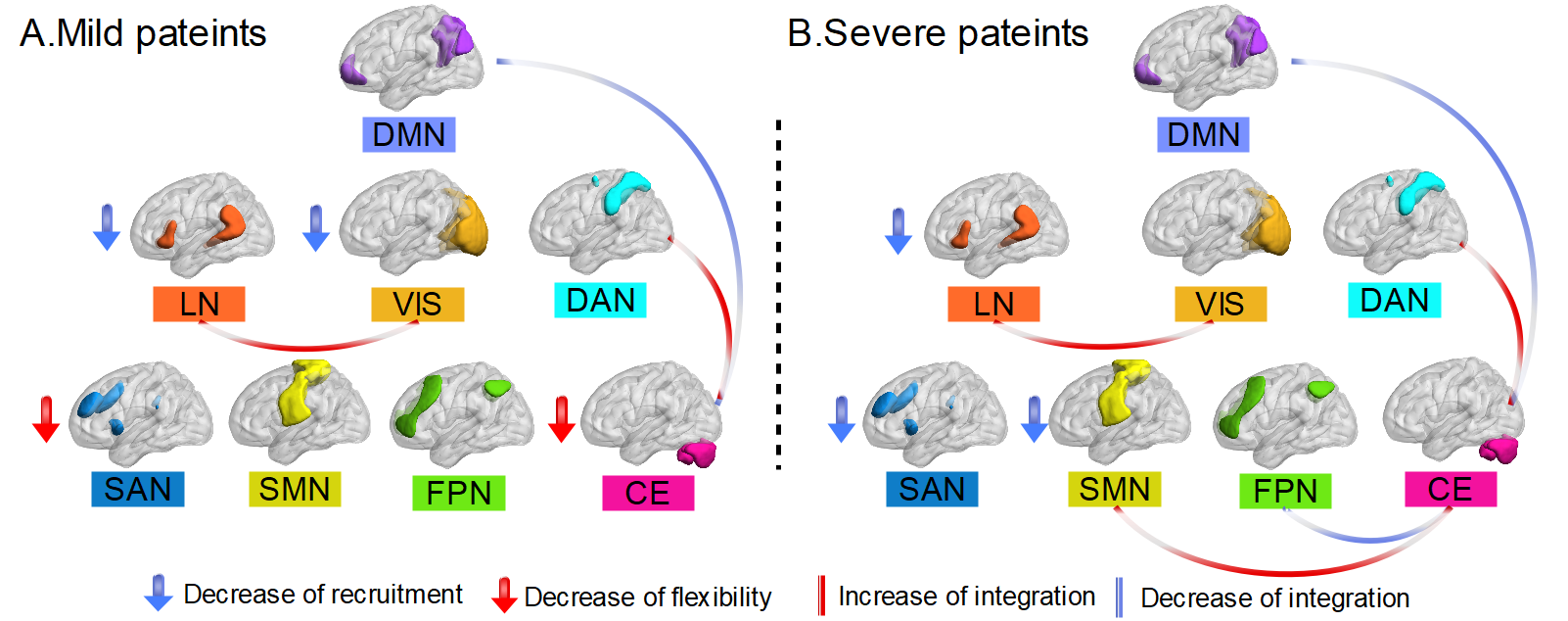}
    \caption{Summary of the reconfiguration patterns of stroke patients with different degrees of symptoms. \textbf(A) Mild. \textbf(B) Severe. \label{fig：summary.png}}
\end{figure*}

\section{Discussion}

As presented, we investigated alterations in the dynamic modules of human brain functional networks across stroke patients with different levels of severity. Specifically, the dynamic functional network changes were modelled across three groups of patients: healthy, mild stroke, and severe stroke, by using a multilayer network method. According to the inherent dynamics of the brain in a resting state, the post-stroke multilayer networks were constructed, from which three measures (recruitment, integration, flexibility) were derived. These three dynamic network measures characterize the brain network reconfiguration after a stroke from different points of view. Given the trends observed in these measures across the three states of patients, we can learn that mild and severe patients exhibit different reconfiguration patterns (a summary of the reconfiguration patterns can be seen in \autoref{fig：summary.png}). Mild stroke patients can be summarized as having a reduction in recruitment in VIS and LN, decreased DMN-CE and increased VIS-LN and DAN-CE integration, and declined SAN and CE flexibility. In contrast, severe patients were characterized by reduced recruitment in SMN, SAN, and LN. In addition to the same integration trend as in mild patients, severe patients were also observed to additionally have raised SMN-CE and lower FPN-CE integration. To the best of our knowledge, this is the first study applying a multilayer network model and evaluating multiple dynamic measures to explore the dynamic reconfiguration of the functional brain network following a stroke. 
We believe these findings could underpin post-stroke functional plasticity and reorganization and may enable new insight into rehabilitation strategies to promote recovery of function.

\subsection{Whole Brain Static Modularity Across Stroke Patients with Different Levels of Severity} 

Using an edge-density approach, lower modularity in post-stroke patients than in healthy controls was observed. This result was consistent with previous studies providing evidence that modularity in resting-state post-stroke patients is reduced~\cite{RN139, RN18}. While a significant group effect in modularity was not observed between the severe and mild patients, much lower modularity was detectable in the mild patients than in the severe patients compared to controls. Despite there being no direct evidence proving that the relationship between modularity and post-stroke severity fits a U-shape of the curve, such a plausible relationship has been depicted in previous dynamic functional connectivity analyses for acute stroke patients. For example, mild patients prefer to stay in a densely connected brain state characterized by a lower level of modularity than severe patients~\cite{RN9}. This improvement of rich modularity in patients with high severity is not exclusive to stroke disease but can also be observed in Parkinson's disease~\cite{RN130} and traumatic brain injury~\cite{RN262}. However, post-stroke recovery studies also present a linear relationship between modularity and behaviors: modularity continually increases as the severity of clinical symptoms alleviates, until it reaches a normal level. Duncan et al.~\cite{RN19} reported that aphasia patients with improved narrative production following therapy had increased modularity in resting state networks. A recent study of large-scale stroke patients by Siegel et al.~\cite{RN139} demonstrated that two weeks after stroke, patients’ functional deficits had been alleviated, and in parallel with this function recovery, the modular structure reemerged and was enriched. Theses distinctly different trends in modularity can be explained by differences between the between-person and the within-person effects. Within-person effects emphasize the trend over a certain period for a specific group, while between-person effects fuse multiple differences that the groups exhibit~\cite{RN273}.

On the other hand, the reduction of static modularity can also be interpreted as decreased segregation between functional domains or networks. The human brain can be parcelled into various functional domains. Functional segregation refers to the independent processing ability of the locally isolated domain to define specific brain functions (when it comes to cooperation between the distributed domains, it refers to functional integration). Although the relationship between modularity and segregation is unclear, higher modularity generally suggests greater segregation~\cite{RN9, RN264}. Hence, from the reduced modularity observed in our results, lower segregation in stroke patients can be inferred. This finding is consistent with previous studies that interpreted the lower post-stroke modularity as a result of decreasing network segregation~\cite{RN19, RN139, RN3}. Results from dynamic functional connectivity analysis also implied that severe patients prefer a state featuring a higher level of segregation between motor domains~\cite{RN260, RN9}. Nevertheless, inferring domain segregation trends from static modularity should be taken with caution. When we link modularity and segregation within patient subgroups, a non-linear relationship might exist. There was a hypothesis related to post-stroke aphasia posited by Duncan et al.~\cite{RN19} that modularity and segregation might follow an inverse U-shaped curve. This is contrary to the inference that severe patients have higher segregation than mild patients because of the lower modularity.

Collectively, whole brain modularity is significantly reduced after stroke. This alteration fits a U-shape curve as severity increases, but between-group effects dilute this result. Besides, modularity should not be used as a biomarker of functional segregation. Post-stroke alteration of functional segregation should be supported by extra observations.

\subsection{Functional Segregation and Integration from Multilayer Dynamic Network Measures}

The brain modular organization is not static but instead fluctuates constantly in response to brain functional demands, even in a resting state. In light of that fact, it is more reasonable to conclude post-stroke functional segregation and integration trends from the dynamic measures than from the static modularity. 

First, there was reduced recruitment within functional networks in both mild and severe patients, indicating that the regions within these networks orchestrate each other less often after stroke. It suggests that VIS and LN in mild patients, SMN, SAN and LN in severe patients tend to process information in an isolated state. Given the bodily functional deficits related to these functional domains, this isolation might correlate with the specific severity of clinical symptoms. Particularly, severe patients were found to have significantly lower recruitment in SAN than mild patients, suggesting that patients with higher clinical symptoms have much lower SAN segregation. The inverse U-shaped curve observed complements the work of Duncan et al.~\cite{RN139}. This pattern cannot be seen in the static modularity analysis, as this recruitment calculation was embedded with temporal information.

Next, regions with significant group differences show an increasing trend of integration in severe patients. Recall that the definition of nodal-level integration indicates that these regions in post-stroke patients tend to interact more with other modules beyond the predefined function domain. Regarding the alteration in the integration of specific domains, there was no significant difference observed. However, the pairwise integration between functional domains has been significantly altered. Regardless of patient groups, increased integration between VIS and LN and between DAN and CE, and decreased integration between DMN and CE can be observed. The post-stroke integration changes suggest that the stroke lesions alter the information transfer between domains instead of within the domain. Besides, it is worth noting that mild patients show less integration alteration than severe patients. In addition to the shared changes with mild patients, integration in severe patients between FPN and CE was decreased, and between SMN and CE was increased. This alteration suggests a link between the interaction between-domain and the level of post-stroke clinical symptoms. Interestingly, the between-network interaction alteration follows a specific balancing mechanism: some pairwise integrations increase while others decrease.

Collectively, stroke groups with different severity levels express similar dynamic patterns. For either the mild or severe patients, the recruitment and integration trend suggests a trade-off between network segregation and integration: segregation increases between some systems, and integration increases or decreases between others. Previous studies have shown that the dynamic measure which has been affected, e.g. by consciousness or by neurological disorders like temporal lobe epilepsy~\cite{RN26}, misdiagnosis of bipolar disorder~\cite{RN266} showed an evident recovery trend. As the exact quantitative relationship between the reduced alteration in integration and stroke recovery was not calculated, it is possible to conclude that post-stroke recovery involved integration and recruitment normalization.

\subsection{Dynamic Functional Network Analysis and Network Flexibility}
Dynamic functional network analysis (DFNC) has recently become popular when working with resting-state fMRI~\cite{RN256, RN267, RN8}. Given the capacity of DFNC to delineate spontaneous variation of functional connectivity, a growing number of such methods have been applied in various scenarios that need to assess high-level network flexibility, not only stroke~\cite{RN260, RN9, RN225, RN136, RN62, RN205} but also Parkinson’s disease~\cite{RN130}, Huntington’s disease~\cite{RN204}, migraine~\cite{RN268} and normal brain ageing~\cite{RN269}. However, the definition of flexibility and the conditions for it may differ across studies. For example, Bonkhoff et al.~\cite{RN9} defined the transition of transient states as network flexibility and reported that the patient groups with motor impairment prefer to shift states when compared to controls. This conclusion was generalized to the broader stroke population in their subsequent work~\cite{RN260}. Besides, the temporal variation of functional connectivity can be seen as a metric measuring network flexibility fluctuation. Chen et al.~\cite{RN136} demonstrate that FC variability is higher in stroke patients than in healthy controls. A similar increase in temporal variability in the ipsilesional precentral gyrus at the subacute stage was reported by Hu et al.~\cite{RN205}.

As a measure that is also developed on the time-varying network in the same way as the flexibility described above, the flexibility used in this study reflects a network's allegiance to a predefined functional domain. Hence, network flexibility here emphasises the temporal variations in the network configuration. The higher the flexibility, the more frequently the network engages in between network interactions. From the results, SAN and CE's flexibility in mild patients was significantly lower than in controls and severe patients. The SAN plays a crucial role in identifying the most biologically and cognitively relevant events for adaptively guiding attention and behaviour and constitutes a critical interface for cognitive, homeostatic, motivational, and affective systems. Besides, the extensive connections with the cerebellum (such as the basal ganglia and thalamus) are linked to perceptual, cognitive, and motor processes. The lower flexibility found in mild patients indicates that the role of two networks in brain communication is decreasing, which explains the post-stroke cognitive deficit~\cite{RN270}. As there are no networks with significantly different flexibility in severe patients, we can make no conclusions regarding the flexibility level being dependent on the severity level. 
 
Nevertheless, network flexibility has been found to be associated with verbal creativity~\cite{RN276}, attention~\cite{RN275}, fatigue~\cite{RN278}, depression~\cite{RN274}, and high-order cognitive functions~\cite{RN277}. These studies confirm the neurobiological basis of network flexibility during adaptive brain processes. Hence, it is natural to speculate that flexibility is positively correlated with stroke severity. This alleged relationship supports a post-stroke neuron bypass theory~\cite{RN134}, i.e., through network reorganization, the neurons bypass the brain regions with deficits and attempt to form new connectivity. These newly forming pathways drive the switching rate to fluctuate wildly to an optimal connectivity pattern. However, if the flexibility could foresee the exacerbation or improvement of brain function after a stroke still needs to be verified.~\cite{RN263}. In the future, it is worth trying to acquire long-term post-stroke behavioural markers to investigate the link between flexibility and brain function.

\section{Conclusion}
In this study, a multi-layer network analysis-based method was proposed to study the dynamic changes in the brain of stroke patients with different severity levels. The indistinguishable network reconstruction pattern with severity dependencies demonstrates the potential of this dynamic method in capturing key features of clinical symptoms of a stroke. Patients with severe deficiencies tend to reduce recruitment and increase integration between networks. However, patients with mild defects have lower network flexibility. These observations provide clear evidence for brain network reconstruction after stroke, which static modular methods cannot do. Therefore, this study expands the resting state fMRI-based functional connectivity analysis methods applied for post-stroke patients. It is worth noting that the degree of functional impairment after stroke seems to be related to differences in dynamic network reorganization patterns among stroke patients. In clinical practice, these findings could help observe the transition from a severe to a mild state during stroke patients’ rehabilitation process. Moreover, the proposed dynamic method could assist clinicians in performing accurate prognosis assessments or could be used as a brain status monitoring method while conducting the therapeutic intervention.

\bibliography{reconfiguration}
\end{document}